\documentclass[a4paper,conference]{IEEEtran}
\IEEEoverridecommandlockouts
\usepackage{amsmath,amssymb,amsfonts}
\usepackage{algorithmic}
\usepackage[noadjust]{cite}
\usepackage{microtype}
\usepackage{nomencl}
\usepackage{eurosym}
\usepackage{graphicx}
\usepackage{soul}
\usepackage{textcomp}
\usepackage{siunitx}
\usepackage{xcolor}
\usepackage{microtype}
\usepackage{hyperref}
\usepackage[noabbrev,capitalise]{cleveref}
\crefname{subsection}{Subsection}{Subsections}
\crefname{figure}{Fig.}{Fig.}

\begin{document}
    \title{Online Network-Constrained Dispatch of Distributed Generators in Radial Networks
        \thanks{This project has received funding from the European Union's Horizon 2020 research and innovation programme under Marie Sklodowska-Curie grant agreement No.\ 675318 (INCITE).}
    }
    
    \author{\IEEEauthorblockN{Hazem A. Abdelghany}
    \IEEEauthorblockA{\textit{Electrical Engineering,} \\
    \textit{Mathematics and Computer Science} \\
    \textit{Delft University of Technology}\\
    Delft, The Netherlands \\
     Electrical and Control Engineering,\\Arab Academy for Science, Technology,\\and Maritime Transport\\ Cairo, Egypt}
    \and
    \IEEEauthorblockN{Carlos Ocampo-Martinez}
    \IEEEauthorblockA{\textit{Institut de Rob\`otica i Inform\`atica Industrial} \\
    \textit{Universitat Polit\`ecnica de Catalunya}\\
    Barcelona, Spain \\
    carlos.ocampo@upc.edu}
    \and
    \IEEEauthorblockN{Nicanor Quijano}
    \IEEEauthorblockA{\textit{Electrical and Electronics Engineering }\\
    \textit{Universidad de los Andes}\\
    Bogot\'a, Colombia \\
    nquijano@uniandes.edu.co}
    }

    \maketitle
    
    \begin{abstract}
    High penetration of distributed generation will be characteristic to future distribution networks. The dynamic, intermittent, uncertain and deregulated nature of distributed generation raises the need for online, distributed economic dispatch techniques. In this paper, we demonstrate the application of such approaches using population dynamics. We propose a congestion management algorithm and demonstrate the notable properties and requirements of the proposed approach.
    \end{abstract}    
    \section{Introduction}
The power industry is undergoing a transition due to developments in renewable energy and the deregulation of power systems. Consequently, future power systems will be characterized by high penetration of small, local distributed generators (DGs). Optimal operation of the distribution system requires optimal dispatch of DGs such that total generation cost is minimized while supply/demand matching, generation limits, and network constrains are maintained. However, the dynamic, intermittent, uncertain and deregulated nature of distributed generation makes this operation a complex task. When conventional dispatch methods (i.e.,\ centralized and offline) become ineffective, the need arises for online and distributed methods for dispatch of distributed generation.

Distributed economic dispatch (ED) is essentially a distributed resource allocation problem. Conventional distributed optimization approaches such as dual decomposition (DD), Lagrangian relaxation, distributed gradient algorithms and the alternative direction method of multipliers (ADMM) have been used to solve the problem in~\cite{conv_1,conv_2,conv_3,conv_4}. In~\cite{PD_comm}, a comparison is made between distributed population dynamics and conventional methods (i.e.,\ DD and ADMM). In general, distributed population dynamics require less information and communication when compared to DD and ADMM. This follows on the introduction of population dynamics for the dispatch of distributed generation in~\cite{PD} and the introduction of distributed population dynamics for optimization and control applications in~\cite{DPD}. 

In this paper, we consider radial distribution networks supplied by distributed generators. We propose an online distributed ED mechanism that takes into account network flow constraints. An online dispatch mechanism can respond to either changes in demand or changes in the availability of DGs. The use of distributed algorithms enables multiple agents to reach a global objective without the need for a centralized optimizer, thus introducing scalability, a degree of privacy, and opennes to heterogeneity.

In resource allocation, replicator dynamics have been proposed in~\cite{PD} and extended for the case with local information constraints in \cite{EGTMain}. In this case, information exchange in the game is represented by a graph where vertices represent the different strategies and edges represent the ability to exchange information between them~\cite{EGTMain}. In~\cite{EGTMain} one of the main applications, ED of distributed generation, was demonstrated. However, only generation limits were considered. An extension was made in~\cite{Nikos} where network flow limits were considered. The authors proposed an algorithm that implements replicator dynamics for ED and proposed a network overflow mitigation approach.

Although this work addresses the same issue, we differentiate between this work and previous works in two aspects:i) We propose the use of the Smith dynamics. This avoids cases in~\cite{Nikos} where a strategy becomes extinct while using replicator dynamics. ii) We propose a novel congestion management algorithm. Unlike the approach used in~\cite{Nikos}, the proposed approach is embedded within the fitness calculation and does not require additional runs of the population game.

In the following \cref{sec: Methodology}, we formulate the ED problem and propose the solution approach including a novel congestion management algorithm. We give a background on the concepts of population dynamics, their most notable properties, and their suitability for distributed resource allocation. We also highlight the differences between this work and the most related work in~\cite{EGTMain,Nikos}. We demonstrate the performance of the control approach in \cref{sec: Results}. For simulation purposes, we use a modified version of the European low voltage test feeder~\cite{ELV}. The simulations aim to show the ability of the proposed algorithm to converge to the optimal solution in reasonable time. Finally, we conclude by discussing the advantages of the proposed algorithm and highlight future work in \cref{sec: Conclusion}.

\section{Methodology} \label{sec: Methodology}

In this paper, we propose a control strategy for online dispatch of DGs in radial distribution networks. The proposed control strategy accounts for both generator limits and line flow constraints imposed by the physical network.

The objective of the proposed control strategy is to minimize overall generation cost subject to supply/demand matching, generator limits, and line flow constraints. That is,

\begin{subequations}
\label{opt}
\begin{align}
	& \underset{{p_{\text{G}}}_i \forall i \in N}{\min}	&& \sum_{i\in N} C_{i} ({p_{\text{G}}}_i)		\label{eq: obj1},\\
	& \text{s.t.} 			 
	&& {{P_{\text{G}}}_i^\text{min}} \leq {{p_{\text{G}}}_i} \leq {{P_{\text{G}}}_i^\text{max}}	,\;	\forall i \in N
	\label{eq: c1.1},\\
	&&&\sum_{i\in N} {{p_{\text{L}}}_i}=\sum_{i\in N} {{p_{\text{G}}}_i}									\label{eq: c1.2},\\
	&&&\left \vert p_{ij} \right \vert \leq U_{ij}						,\;	\forall i,j \in N 
	,\; (i,j) \in L																		\label{eq: c1.3},
\end{align}
\end{subequations}

where, ${p_{\text{G}}}_i$ is the power generated at node $i$; $N,\,L$ are the sets of nodes and lines in the system respectively; ${{P_{\text{G}}}_i^\text{min}},\,{{P_{\text{G}}}_i^\text{max}}$ are the lower and upper limits of generators; ${p_{\text{L}}}_i$ is the load at node $i$; $p_{ij}$ is the line flow between nodes $i,\,j$; $U_{ij}$ is the line limit for line $(i,\,j) \in L$; $C_{i} ({p_{\text{G}}}_i)$ is the cost of generation for node $i$ represented by
\begin{equation}
    C_{i} ({p_{\text{G}}}_i)=a_i+b_i {p_{\text{G}}}_i+c_i {{p_{\text{G}}}_i}^2, \label{eq:cost}
\end{equation}
where $a_i,\,b_i,\,c_i$ are the coefficients of the generator cost function.

The optimization problem in \eqref{opt} is particularly hard when the system under consideration is subject to changing demand, topology (e.g.,\ mobile generation, electric vehicles, storage), or intermittent distributed generation (e.g.,\ renewables, storage). For a system with such characteristics, an online control strategy is more suitable than an offline one. Also, a distributed control strategy provides a possible approach which achieves scalability and modularity. In our proposed control strategy, generators in the network cooperate to achieve the optimal system-level behaviour.

In our proposed approach, a node controller is located at key nodes in the system. These are nodes hosting a generator or nodes connecting more than two lines (i.e.,\ points where lateral feeders branch from the main feeder. This is the maximum amount of controllers required since, for overflow management, series lines can be represented using the lowest flow limit among them. For any key node, the node controller has three roles:
\begin{itemize}
    \item It measures flow in all lines connected to its node.
    \item It detects and calculates overflow in any of the lines.
    \item If overflow is detected, it broadcasts a penalty/incentive to all nodes electrically connected on its side of the congested line. This is further elaborated in \cref{sec: congestion}.
\end{itemize}
  The node controller of a generator node has an additional role, i.e.,
  \begin{itemize}
      \item It calculates the fitness of the generator node in the population game accounting for the power generated, violation of generator limits, or any broadcast signals received from other nodes indicating congestion.
  \end{itemize}

\subsection{Generator node fitness}
\begin{figure}
    \centering
    \includegraphics[width=0.8\linewidth,keepaspectratio]{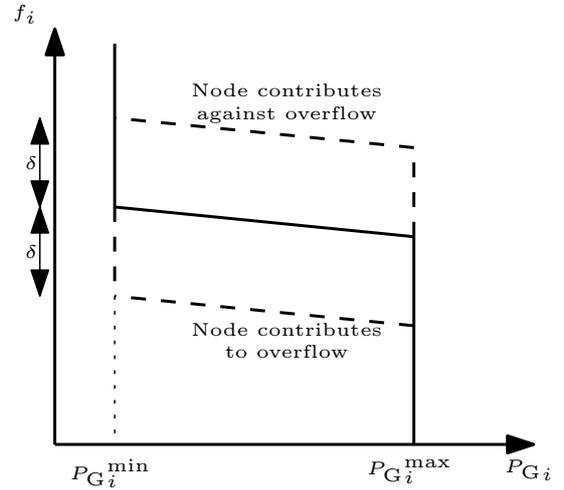}
    \caption{Fitness of a generator node.}
    \label{fig: fitness}
\end{figure}

Assuming a generator cost function for generator $i$ is represented by the polynomial in (\ref{eq:cost}), the ED solution can be obtained when the marginal cost (i.e.,\ derivative of the cost function) of all generators is equal. Therefore, there is a resemblance between the solution of the ED problem and the equilibrium condition of population dynamics~\cite{EGTMain}. The fitness of a generator node can be found by,
\begin{equation}
    f_i=\hat{f}_i+{f_{barr}}_i,
    \label{eq: fitness}
\end{equation}
where $\hat{f}_i$ is the decreasing fitness function within feasible region compensated to yield positive fitness, that is,
\begin{equation}
    \hat{f}_i=B-(b_i+2 c_i {p_{\text{G}}}_i),
\end{equation}
where $B$ is a large positive bias to yield positive fitness values. ${f_{barr}}_i$ is a barrier function that penalizes violation of generation limits found by
\begin{equation}
	{f_{barr}}_i=
    \begin{cases}
        m {p_{\text{G}}}_i				& {p_{\text{G}}}_i < {{P_{\text{G}}}_i^\text{min}}\\
        0								& {{P_{\text{G}}}_i^\text{min}} \leq {p_{\text{G}}}_i \leq {{P_{\text{G}}}_i^\text{max}}\\
        -m {p_{\text{G}}}_i				& {p_{\text{G}}}_i > {{P_{\text{G}}}_i^\text{max}}\\
    \end{cases},
\end{equation}
    where $m$ is the large slope of the barrier function. The fitness function of a typical generator node is shown in \cref{fig: fitness}.

\subsection{Congestion detection} \label{sec: congestion}
\begin{figure}
    \centering
    \includegraphics[width=0.8\linewidth,keepaspectratio]{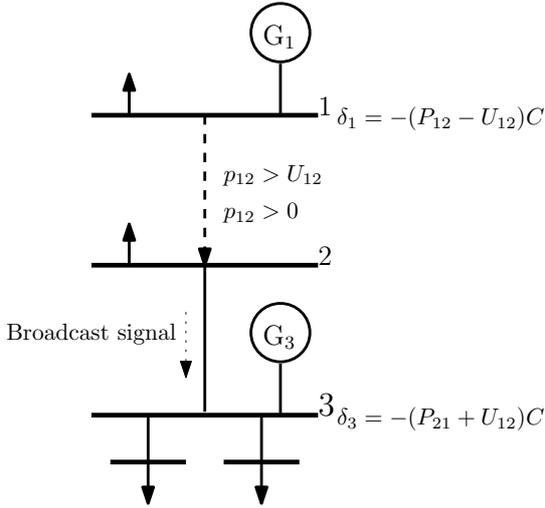}
    \caption{Congestion detection at node 2.}
    \label{fig: congestion}
\end{figure}

Whenever a node controller $i \in N$ detects overflow on a line connected between nodes $i,\,j \in N$, the controller computes a value proportional to the amount of overflow and dependant on the direction of the overflow. A node controller that detects outgoing overflow computes a penalty, while a node controller that detects incoming overflow computes an incentive. The incentive/penalty $\delta_i$ computed by controller of node $i$ is given as,
\begin{equation}
    \delta_i= 
    \begin{cases}
        -(p_{ij}-U_{ij})  C                & p_{ij}>0 ,\, \vert p_{ij} \vert >U_{ij}   \\
        -(p_{ij}+U_{ij})  C                & p_{ij}<0 ,\, \vert p_{ij} \vert >U_{ij}   \\
        0                                       & \text{otherwise}                          \\
    \end{cases},
\end{equation}
where $C\in \mathbb{R}_{>0}$ is a positive constant with a large value. The incentive/penalty signal is then broadcast to all nodes on each side of the congested line (e.g.,\ using power line carrier). The penalty/incentive value is added to the fitness of generator nodes. Therefore, a generator node contributing positively to the overflow is penalized, while a generator node contributing negatively is incentivized. This is illustrated in \cref{fig: fitness}. and, using an academic example, in \cref{fig: congestion}.

Compared to the congestion management algorithm used in~\cite{Nikos}, this approach is embedded within the fitness function calculation of each node. Therefore, the ED algorithm is only required to run once regardless of the existence of line flow limit violations or not. This makes it more suitable for online applications where finding a solution quickly is of high importance.

    \subsection{Population games \& mean dynamics}   
Population games first emerged as a model to represent interactions in large systems~\cite{smith}. They are usually characterized by a large number of agents, who are small, and interact anonymously with a finite number of pure strategies and continuous payoffs~\cite{Sandholm}. Agents in a single population are represented by a continuum of mass with a finite set of pure strategies. The social state represents the ratio of the population adopting each strategy. A payoff function maps each social state a vector of payoffs, one for each strategy. Every population game admits at least one Nash equilibrium~\cite{Sandholm}. Compared to equilibrium play in classical game theory, evolutionary game theory uses evolutionary dynamics to describe how agents adapt (i.e.,\ revise) their strategies in response to their observed strategic environment using a revision protocol. The aggregate behaviour of the society can be described using a mean dynamic. Most notable, replicator dynamics emerge when a population uses imitating revision protocols (i.e.,\ agents adopt the most successful strategies adopted by other agents)~\cite{bauso}. The Smith Dynamics represent pairwise comparison protocols, where agents adopt strategies that will improve their current payoff.

In ED, the population is represented by the total demand in the system, i.e.,\ $P_{d}=\sum_{i\in N} {{p_{\text{L}}}_i}$. Each generator represents a strategy and has a positive fitness decreasing in the ratio of the population that adopts this strategy $x_i=\frac{{p_{\text{G}}}_i}{P_d}$
Replicator dynamics, mainly used in previous literature, are described by
\begin{equation}
    \dot{x}_i=x_i(f_i(x)-\bar{f}(x)) \forall i \in \mathcal{S},
\end{equation}
where $\mathcal{S}$ is the set of strategies, while $\bar{f}(x)$ is the mean fitness of the population at population state $x$, given by
	\begin{equation}
		\bar{f}(x)=\sum_{i \in \mathcal{S}} \frac{{p_{\text{G}}}_i}{P_d} f_i(x).
	\end{equation}
	In ED, this translates to
	\begin{equation} 
		\frac{\dot{{p_{\text{G}}}_i}}{P_d}=\frac{{p_{\text{G}}}_i}{P_d}\left(f_i(x)-\bar{f}(x)\right)\; \forall i \in \mathcal{S}.
	\end{equation}
Replicator dynamics can be rewritten when information exchange is constrained to neighbouring strategies. This is called the local replicator equation \cite{EGTMain}, which is given by
\begin{equation}
    \dot{x}_i=x_i(f_i(x) \sum_{j \in \mathcal{N}_i} x_j - \sum_{j \in \mathcal{N}_i} f_j(x) x_j) \forall i \in \mathcal{S},
\end{equation}
where $\mathcal{N}_i$ is the set of neighbouring strategies to strategy $i$ in the graph representing information exchange. 

It can be seen that using replicator dynamics, a generator with initial population mass $x_i=0$ will always have a change of $\dot{x}_i=0$. This is then called an extinct strategy. The assumption that loads are initially assigned to their nearest generators in ~\cite{Nikos} combined with the use of replicator dynamics lead to cases where an efficient generator will not be utilized due to being assigned no load initially. This results from the extinction property of replicator dynamics (i.e.,\ unused strategies are never subsequently chosen~\cite{Sandholm}). This also means that when using distributed replicator dynamics, if the graph representing the game is not a complete graph, an extinct strategy will lead to truncation of the graph. An in-depth discussion of this issue can be found in~\cite{EGTMain}. This motivates our proposition to use the Smith dynamics. These dynamics can be represented by
\begin{multline}
\dot{x}_i=\sum_{j\in \mathcal{S}}x_j[f_i(x)-f_j(x)]_{+}-x_i \sum_{j\in \mathcal{S}} [f_j(x)-f_i(x)]_{+} \\ \forall i \in \mathcal{S}.
\end{multline}
In~\cite{DSD}, a distributed form of the Smith dynamics was formulated, which can be represented by
\begin{multline}
    \dot{x}_i=\sum_{j\in \mathcal{N}_i}x_j[f_i(x)-f_j(x)]_{+}-x_i \sum_{j\in \mathcal{N}_i} [f_j(x)-f_i(x)]_{+} \\ \forall i \in \mathcal{S}.
\end{multline}
Unlike replicator dynamics, in this case, a strategy with population mass $x_i=0$ can still evolve if it outperforms any of the other strategies adopted by the population. This is more suitable for the case when a more efficient DG has an initial allocation of $0$ (e.g.,\ due to existence of less efficient local generation).

    \section{Simulation \& Results} \label{sec: Results}

\begin{figure}[htbp]
    \centering
    \includegraphics[width=0.8\linewidth,keepaspectratio]{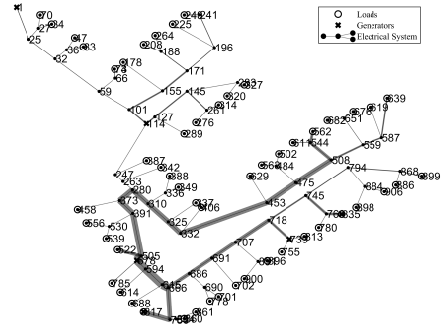}
    \caption{Overview of the modified European low voltage feeder, line weights represent flows at $t=566$.}
    \label{fig: line_flows}
\end{figure}

To demonstrate the effectiveness of the proposed ED approach in use, we implement a modified version of the European low voltage test feeder~\cite{ELV} in MATPOWER and use the population dynamics toolbox to simulate the population game~\cite{toolbox}. While the original European low voltage feeder has 906 nodes, we modify it to represent only key nodes as defined in \cref{sec: Methodology}. The resulting system has 110 nodes. Lines connected in series are replaced by a line with the sum of their impedences. The loads in the system have an aggregate peak of \SI{57}{\kilo \watt}, and vary each minute according to specific load profiles~\cite{ELV}. \cref{fig: line_flows} shows an overview of the modified system with loads highlighted with a circle marker.
Additionally, the original system is supplied solely through the distribution transformer at node 1. For the purpose of the following simulations, we assume five more generator nodes spread randomly with parameters shown in \cref{gen_param}.
\begin{table}[htbp]
\caption{Distributed generator parameters}
\begin{center}
\small\addtolength{\tabcolsep}{-5pt}
\begin{tabular}{|c|c|c|c|c|c|}
\hline
\textbf{Bus}&\multicolumn{5}{|c|}{\textbf{Generator parameters}} \\
\cline{2-6} 
\textbf{no.} & \textbf{$a$(\euro)}&\textbf{$b$(\euro/\si{\mega \watt})}& \textbf{$c$(\euro/\si{\mega \watt^2})} &  \textbf{$P_{\text{min}}$(\si{\kilo \watt})} & \textbf{$P_{\text{max}}$(\si{\kilo \watt})} \\
\hline

1       &  0    &  5    &  0.02 &   0   &           10              \\          
114     &  0    &  1    & 0.01  &   0   &           10              \\
578     &  0    &  3    &  0.025&   0   &           5               \\
739     &  0    &  4    &  0.01 &   0   &           10              \\
817     &  0    &  2.5  & 0.015 &   0   &           20              \\
835     &  0    &  2    & 0.02  &   0   &           10              \\
\hline
\end{tabular}
\label{gen_param}
\end{center}
\end{table}

For the following subsections, we assume that information exchange between different generators is represented by a complete graph. In this case, distributed and centralized population dynamics are equivalent. The discussion on the use of distributed population dynamics is addressed in \cref{sec: Conclusion}. To evaluate our results, we compare the outcome of the population-dynamics-based approaches with the solutions of the DC optimal power flow (DC-OPF) obtained from MATPOWER for the same case. We use the fitness function parameters $B=1000,\,m=400,\,C=1000$. Unless otherwise mentioned, Smith dynamics are adopted for the following results.

\begin{figure*}
    \centering
    \includegraphics[width=\textwidth,keepaspectratio]{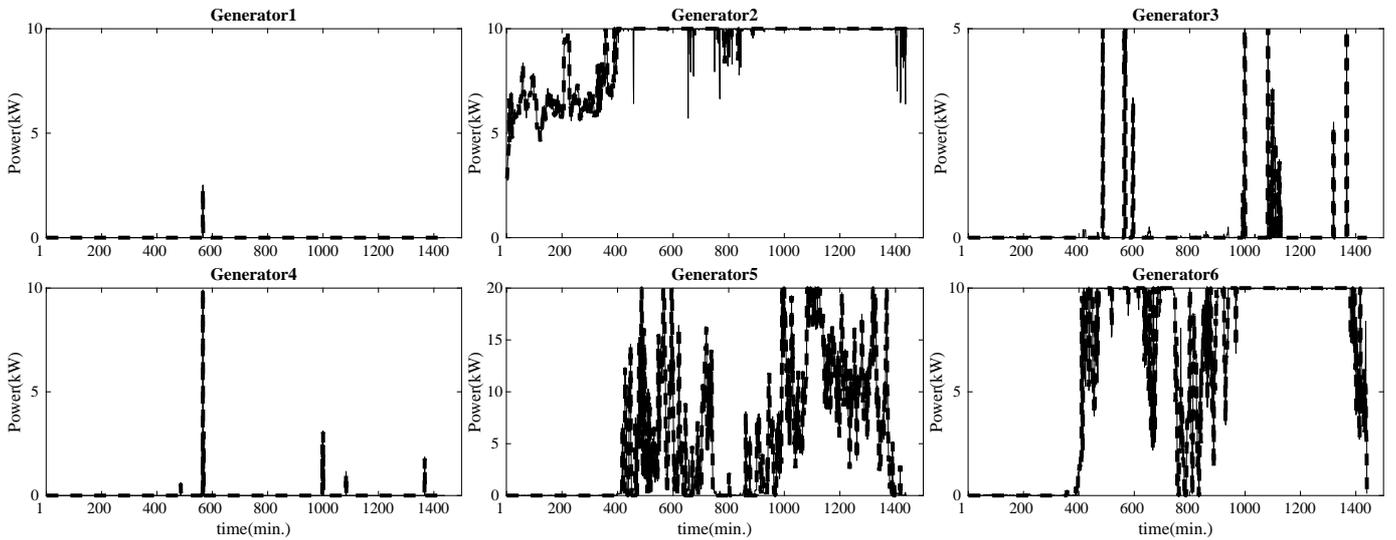}
    \caption{Generator output in base scenario:  Smith dynamics (solid) and DC-OPF (dashed).}
    \label{fig: output_base}
\end{figure*}

For a baseline case, we simulate a complete day. We assume only generator limits and infinite line limits. We show that using the Smith dynamics, generator set points converge to the DC-OPF solution. The outcome of each generator, compared to the optimal dispatch for the whole day, can be seen in \cref{fig: output_base}. \cref{fig: line_flows} shows the line flows represented by the weights of lines at time step 566 of the 1440 minutes in the day. We choose this time-step as it is the time-step with the peak load and therefore the most prone to congestion. Note here that this is the case when no line flow limits are enforced.

\begin{figure}
    \centering
    \includegraphics[width=0.8\linewidth,keepaspectratio]{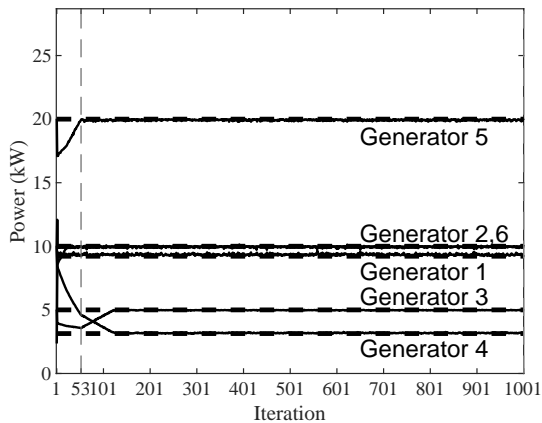}
    \caption{Generator output in congestion scenario: Smith dynamics (solid) and DC-OPF (dashed).}
    \label{fig: congestion_output}
\end{figure}

To demonstrate the effectiveness of the congestion management algorithm, we limit the flow on the lines between buses 505-666 to \SI{28}{\kilo \watt}. We simulate only 1 time-step $t=566$. This leads to an overflow of 4\si{\kilo \watt} to 6\si{\kilo \watt}. \cref{fig: congestion_output} shows the evolution of the generators' set points until congestion is detected in iteration 53 of the population game. Then, Generator 4 (bus 739) is penalized and starts to maintain a set points within its feasible region. Generator 1 (bus 1) increases its set points to compensate for the remaining load while mitigating the congestion detected. The solver uses a time-step of \SI{0.01}{\sec}. Convergence can be observed at iteration 201 (i.e.,\ \SI{2}{\sec}).
   
    \section{Conclusion} \label{sec: Conclusion}
In this paper, we proposed an online distributed dispatch algorithm for generators in radial networks with line flow limits. The proposed approach relies on population games and the implementation of population dynamics to establish the optimal set points. Simulations demonstrate the ability of the dispatch mechanism to converge towards the optimal set points in various cases.

Results assume complete connectivity between different strategies. This means that in this case, both distributed and centralized population dynamics are equivalent. However, the extinction property challenges the implementation of distributed replicator dynamics as they require guarantees that no strategy will be extinct. In ED, this is a strong assumption since some generators may be assigned no load initially. An example of this case can occur when supply provided via the distribution transformer is more efficient than local thermal generation (e.g.,\ at times of low local renewable generation). In this case, load may be assigned to nearby local thermal generation overlooking the more efficient supply option. This is a scenario where replicator dynamics will not converge. Otherwise, distributed population dynamics can achieve the same results obtained by centralized dynamics as shown in previous literature~\cite{EGTMain,PD_comm,DPD,DSD}.
   
This work extends the work previously done in \cite{EGTMain} by including network constraints to a distributed smith dynamics approach. The use of distributed smith dynamics avoids problems inherent in the replicator dynamics due to the extinction property. We propose a novel congestion management algorithm and highlight the difficulties in implementing replicator dynamics. We only consider radial distribution networks as they represent a majority of the systems where such an approach is needed. However, generalizing the same concept for a general form of distribution network is an interesting and challenging task that remains for future work. 
    \small
\bibliographystyle{IEEEtran}


\end{document}